\documentclass[final,5p,times,twocolumn]{elsarticle}

\usepackage{lineno}

\biboptions{sort&compress}

\usepackage{esvect}
\usepackage{textgreek}
\usepackage{textcomp}
\usepackage{amssymb}
\usepackage[export]{adjustbox}
\usepackage{subcaption}
\usepackage{caption}
\usepackage{wrapfig}
\usepackage{graphicx}
\usepackage{lipsum}
\usepackage{float}
\usepackage{bm}
\usepackage{makecell}
\usepackage{siunitx}
\usepackage{amsmath}
\usepackage{xcolor}
\usepackage{tabularx}
\usepackage{tikz}
\journal{Acta Materialia}

\begin{document}
\begin{frontmatter}

\title{Examination of computed aluminum grain boundary structures and interface energies that span the 5D space of crystallographic character}

\cortext[cor1]{Corresponding author}
\author[label1]{Eric R. Homer\corref{cor1}}
\ead{eric.homer@byu.edu}
\author[label2]{Gus L. W. Hart}
\author[label2]{C. Braxton Owens}
\author[label2]{Derek Hensley}
\author[label2]{Jay Spendlove}
\author[label2]{Lydia Harris Serafin}

\address[label1]{Department of Mechanical Engineering, Brigham Young University, Provo, UT, 84602, USA.}
\address[label2]{Department of Physics and Astronomy, Brigham Young University, Provo, UT, 84602, USA.}

\begin{abstract}
The space of possible grain boundary structures is vast, with 5 macroscopic, crystallographic degrees of freedom that define the character of a grain boundary. While numerous datasets of grain boundaries have examined this space in part or in full, we present a computed dataset of over 7304 unique aluminum grain boundaries in the 5D crystallographic space. Our sampling also includes a range of possible microscopic, atomic configurations for each unique 5D crystallographic structure, which total over 43 million structures. We present an overview of the methods used to generate this dataset, an initial examination of the energy trends that follow the Read-Shockley relationship, hints at trends throughout the 5D space, variations in GB energy when non-minimum energy structures are examined, and insights gained in machine learning of grain boundary energy structure-property relationships. This dataset, which is available for download, has great potential for insight into GB structure-property relationships.
\end{abstract}


\end{frontmatter}


\section{Introduction}
The interfaces or grain boundaries (GBs) separating the individual crystals in polycrystalline materials can have a significant effect on numerous material properties. In fact, in the 1990s, a field emerged called GB engineering that demonstrated large improvements in properties by controlling the population of different grain boundary types \cite{Palumbo:1998ux,Watanabe:2009jm,Randle:2010jw}. Notable achievements in GBE include a decreased intergranular corrosion susceptibility \cite{Lin:1995do}, a gain in cycle life of lead acid batteries \cite{Lehockey:1999tt}, a decrease in creep rate \cite{Thaveeprungsriporn.1997.MMTA}, an increased strength, ductility, fracture toughness \cite{Bechtle.2009.ActaMaterialia},  among others. Unfortunately, many of these improvements were limited to low stacking fault energy FCC materials that readily form twins. To enable GBE generally we need more information about the range of energies and structures that are possible in different GB types. In other words, we need a more complete understanding of GB structure-property relationships over the full range of GB types.

Grain boundary structures are defined by both their macroscopic and microscopic degrees of freedom. The macroscopic degrees of freedom of a grain boundary are defined by five crystallographic parameters, often referred to as the GB character. Three parameters describe the misorientation between two grains and two parameters define the orientation of the boundary plane (BP) separating the two grains. The microscopic degrees of freedom of a GB are defined at the most basic level by the positions of each atom, where $N$ atoms would have $3N$ degrees of freedom. Various metrics have been developed to describe the atomic structure of GBs in simpler terms, including the structural unit model \cite{Balluffi:1984kk,Sutton:1989vz,Han:2017io}, polyhedral unit model \cite{Banadaki:2017dk}, and more recently by various collections of local atomic environments \cite{Tamura:2017ja,Gomberg:2017ha,Zhu:2017vna,Rosenbrock:2017gl,Sharp:2018,Rosenbrock:2018ug,Priedeman:2018gm,Snow.2019.StructureML,Patala:2019,Homer:2019fz,Zapiain:2020}.

It must also be noted that for every unique macroscopic structure description, there are a multiplicity of microscopic degrees of freedom where the atoms are arranged differently \cite{10.1007/978-1-4757-0181-4_6,  Frolov:2013cq,Han:2016fi,Hickman:2017fd,Zhu:2017vna,Meiners:2020fd}. These various configurations are typically referred to as metastable configurations or GB phases. In experiments, these can be seen as varying atomic configurations along a single GB \cite{Meiners:2020fd}. In modeling and simulation, most research focuses on the minimum energy configurations because nature would drive toward that state. However, a statistical mechanical approach using these metastable states can be used to predict finite temperature equilibrium and non-equilibrium properties \cite{Han:2016fi}.

While the 5 macroscopic degrees of freedom are significantly fewer than the $3N$ microscopic degrees of freedom, especially considering the metastable states, they still constitute a large space that is challenging to fully resolve. Symmetries are present in this 5D space that reduce the amount of data required to explore it \cite{Patala:2013jj}, but these symmetries can be as problematic as they are helpful because researchers do not always use the unique descriptions of GBs, such as the disorientation\footnote{A disorientation provides a single unique description for many symmetrically equivalent misorientations and is defined by the minimum misorientation angle when the misorientation axis is found in the standard stereographic triangle.} or a unique definition of the BP, such as described in \cite{Homer:2015ie}. 

In any case, this space is sufficiently vast that historically, these 5 degrees of freedom have frequently been simplified down to one degree of freedom, completely ignoring BP and disorientation axis, focusing only on the disorientation angle. This is often simplified even further into low vs.\ high disorientation angles or special vs.\ not special, where special GBs are comprised of low $\Sigma$-value coincidence site lattices (CSLs) and low disorientation angle boundaries. While these simplifications have their uses, some of which will be reinforced by this work, they still represent an incomplete characterization of the GB.

There have been many notable investigations of GB structure-property relationships over numerous different properties, but it is beyond the scope of this work to provide a full review of these efforts beyond highlighting some important works \cite{Gleiter:1982km,Wolf:1992we,Sutton:1995ux,Gottstein:2010wy,Rohrer:2011kp}. However, we do wish to highlight those studies that have either been particularly insightful or unique in their investigation of GB interface energy over some portion, if not all, of the 5D macroscopic GB space. Perhaps the most well-known of these relationships is the Read-Shockley relationship, which predicts GB energy and atomic structure as a function of disorientation angle, in the most basic case for tilt or symmetric tilt GBs \cite{Read:1950um}. This relationship has proven remarkably useful over the years and will see reinforcement in this work. Other noteworthy models include the broken bond GB energy model by Wolf \cite{Wolf:1991wh,Wolf:1990ud}, the Frank-Bilby equation to relate microscopic and macroscopic degrees of freedom \cite{Cahn:2006ku,Yang:2010bs,Sangghaleh:2018ij} and the Wulff construction and Cahn-Hoffman capillarity vector to understand the role of BP \cite{Hoffman:1972vi,Cahn:1974tc,Wheeler:1999tx,Balluffi:2005tw}. More recently Bulatov, Reed, and Kumar created a GB energy function for FCC materials that predicts energy for four different FCC metals based upon the 5 macroscopic degrees of freedom \cite{Bulatov:2014bz}.

Of course all these models require data upon which to learn and predict. Numerous model, experimental, and computed datasets have been developed to provide insight into GB energy structure-property relationships. Older datasets relied on hard sphere approximations \cite{Frost:1982vc} while newer datasets take advantage of advances in characterization and computation. For example, advances in computing have also allowed larger and more detailed atomic structures of GBs to be simulated. Recent experimental datasets take advantage of automated serial sectioning technologies \cite{Spowart:2006:AutomatedSerialSectioning,Dillon:2009ec,Echlin:2020:SerialSectioning} to recover all 5 macroscopic degrees of freedom and morphology of polycrystalline microstructures.

While datasets of computed GBs date back several decades \cite{Wolf:1989kv,Wolf:1989ts,Wolf:1990fk,Wolf:1990um}, a dataset created by Olmsted, Foiles, and Holm \cite{Olmsted:2009ge} has received particular attention because it was the largest for its time and was made available to numerous researchers. The dataset consists of 388 GBs of Ni and Al and their corresponding minimum energies and atomic structure configurations. The number of GBs was constrained by all CSL configurations that could fit within an orthogonal periodic BP of a specific size using rational indices. The 388 GBs were each unique in terms of the 5D macroscopic configuration, though the sampling is not uniform across the space. There are 72 unique disorientations in the dataset and some disorientations sample many BPs, such as the $\Sigma3$, while others sample as few as 2 BPs. The atomic structures were created by sampling a number of construction degrees of freedom, such as translation of one grain relative to the other, and using molecular statics conjugate gradient minimization to find the minimum energy structure for each. Embedded-atom-method (EAM) potentials were used to simulate the atomic interactions. Only the minimum energy structure and energy were saved and reported and all other metastable configurations were discarded.

While Olmsted et al. conclude that disorientation angle is insufficient to determine the GB energy, specific twist GBs were shown to follow the Read-Shockley relationship. The overall trend is generally Read-Shockley like, but the BP and disorientation axis clearly play important roles that cannot be ignored. Furthermore, they find that GB energy correlates with excess volume at the GB \cite{Olmsted:2009ge}.

Numerous other computed GB datasets have since been created. Ratanaphan, Olmsted et al.\, created a dataset similar to Olmsted's FCC GBs that consisted of 408 GBs in BCC Fe and Mo \cite{Ratanaphan:2015ht}. Tschopp et al.\ created a dataset of 174 symmetric and asymmetric tilt GBs in both Al and Cu \cite{Tschopp:2015bv}, which are available for download \cite{Tschopp:2015:GBdataset}. Priedeman et al.\ created a dataset of 126 $[1\,0\,0]$ symmetric tilt grain boundaries \cite{Priedeman:2018gm} and Erickson and Homer created a dataset of 346 GBs, all with $[1\,0\,0]$ disorientation axes \cite{EricksonHomer:2020hd}, both of which are available for download \cite{Homer:2020:100GBdataset}. Guziewski et al.\ modified a Monte Carlo GB optimization approach by Banadaki and Patala \cite{banadaki:2018} to create a dataset of 344 grain boundaries in Si and SiC \cite{guziewski:2020}. It is important to note that the Monte Carlo approach used by Guziewski et al.\ allowed them to keep large datasets of metastable structures as well. Kim et al.\ used two approaches to sample general and special GBs \cite{kim:2011}. To examine general GBs they produced 66,339 evenly distributed Fe grain boundaries in the 5D space. The special GBs were examined by choosing 26 CSLs and generating 2366 evenly spaced GBs in this space. It should be noted however that this approach does not use the typical periodic boundary conditions and only employs a single initial condition, so it is not clear how the measured GB energy relates to the minimum GB energy each boundary might be capable of achieving. YongFeng et al.\ recently generated 230 GBs in UO\textsubscript{2} and CeO\textsubscript{2} \cite{YongFeng:2022:GBenergyUO2}. Finally, a recent GB dataset by Zheng et al.\ was created using density functional theory (DFT) \cite{Zheng:2020}, whereas most datasets are generated using emperical potentials. In this dataset, available through the Materials Project \cite{Jain:2013ku}, 327 GBs were created over 58 different elements, with 10 GB types for fcc and bcc and one GB type for hcp.

Experimental datasets of GBs that recover all 5 macroscopic degrees of freedom often infer GB energy values based on population, geometry of the GB network, or thermal grooving. For example, one can utilize the Herring equation or Cahn-Hoffman capillarity vector to reconstruct the GB energy based on the geometry of triple junctions, where three GBs come together and an energy balance of the three GBs would allow the triple junction to achieve equilibrium \cite{Adams:1999vq,Morawiec:2000vb,Gottstein:2010wy}. One can also infer GB energy based on GB populations, where it is assumed that low energy GBs would show up with high frequency and high energy GBs would show up with low frequency \cite{Saylor:2003gm}. Using one of these methods, GB energy has been, or can be, inferred in the following experimentally obtained GB datasets. These include measurements in Ni \cite{Li:2009rg}, Al \cite{Saylor:2004bp,Holm:2011hj}, NiAl \cite{Mishin:2005:NiAl_GBenergy}, ferritic steel \cite{Beladi:2013,Zhong:2017vd}, GB engineered Ni and Cu \cite{Randle:2008ig}, yttria \cite{Dillon:2009ec}, and magnesia \cite{Saylor:2003gm}, among others \cite{Dillon:2009:SerialSectionGBenergy}, of which many of these datasets are available for download \cite{Rohrer:GBdataset}.

In direct comparisons, Amouyal et al.\ find good agreement between measured and computed GB energies in NiAl \cite{Mishin:2005:NiAl_GBenergy}. Using Ni datasets, Rohrer et al. performed a large-scale comparison between measured and computed grain boundary energies \cite{Rohrer:2010df}. The work concluded that the experimental and computational results validate each other for boundaries that are appropriately represented in both data sets. In later work, Holm et al.\ demonstrate that that one can also validate GB energy calculations using the grain boundary character distribution \cite{Holm:2011hj}.

GB datasets are also crucial to the Materials Genome Initiative focus to develop new materials on a shorter time line, because  digital data is identified as one of the three pillars in the materials innovation infrastructure, along with computational and experimental tools  \cite{NationalScienceandTechnologyCouncil:2011ub}. As will be illustrated below, the Olmsted GB dataset is an excellent example of how digital data has been used in numerous studies to examine a variety of phenomena \cite{Olmsted:2009ge}. 

The FCC GB energy function mentioned earlier, created by Bulatov et al., was trained on the Olmsted dataset and showed that four FCC metals all had similar GB energy trends across the 5D space \cite{Bulatov:2014bz}. Furthermore, the energy prediction could be changed between all four FCC metals by simply changing two parameters related to the twin and random GB energy values. 

Olmsted's dataset was also used to gain insight into energy variation through the 5D space. Bulatov et al. found interesting properties about the energy cusps \cite{Bulatov:2014bz}. Homer and Patala built upon Patala's work on the symmetries of BP orientations \cite{Patala.2013.PhilosophicalMagazine} to identify fundamental zones (or irreducible spaces) in the BP degrees of freedom \cite{Homer:2015ie}. Using these BP fundamental zones, they found that GB energy in the Olmsted dataset varied smoothly for similar BP orientations of the same disorientation as a function of disorientation angle. This trend was later confirmed more generally by Erickson and Homer in their dataset of 346 $[1\,0\,0]$ disorientation axis GBs \cite{EricksonHomer:2020hd}.

Olmsted's dataset has also been used to examine GB mobility \cite{Homer:2014hr,Yu:2019jq}, GB shear coupling \cite{Homer:2013ce,Yu.2021.Materialia}, excess volume \cite{Pal:Deng:2021:spectrumexcessvolume}, and others
\cite{Snow.2019.StructureML,Johnson:2021fj,Baird:5DOFinterp}. More recently Olmsted's dataset of atomic structures were used as input to predict GB energy by machine learning \cite{Rosenbrock:2017gl,Rosenbrock:2018ug,Homer:2019fz}. Priedeman et al.\ built on those methods and used dimensionality reduction techniques to illustrate a connection between the macroscopic and microscopic degrees of freedom \cite{Priedeman:2018gm}.

The other GB datasets have been used for numerous purposes. For example, Zheng developed an improved predictive model for the GB energy of different elements based on the cohesive energy and shear modulus using the DFT dataset mentioned previously \cite{Zheng:2020}.  Tamura et al.\ also developed a predictive model using machine learning, where accurate predictions are reported from training sets with as few as 10 GBs \cite{Tamura:2017ja}.

It is clear that GB datasets have tremendous value, and that, to continue the efforts of GB engineering, more complete structure-property relationships are required. But to obtain more complete relationships  the datasets must span the full range of macroscopic degrees of freedom. Additionally, it would be beneficial if the datasets also provide insight into the range of microscopic degrees of freedom.

In this work, we present a computed dataset of aluminum GBs that span the entire 5D space and provide insight into the range of metastable atomic configurations. In total, 7304 unique aluminum GB structures are examined, with a full set of metastable atomic structure configurations for each. All together, over 43 million GB structures and their respective energies are represented in this dataset. In this work we describe the methods used to construct the dataset, examine general statistics of the dataset, illustrate GB energy trends in subspaces of the 5D GB character, illustrate trends in energy distributions of the metastable configurations, and examine basic machine learning predictions using the atomic structures as input. Finally, the dataset will be made available so that it can serve as a tool of understanding to the entire scientific community \cite{Homer:2022:AlGBdataset}.

\section{Methods}

\subsection{GB selection procedure}


As noted above, the macroscopic (crystallographic) character of a GB is defined by 5 parameters, 3 for the disorientation and 2 for the boundary plane. In our approach to generating our dataset, we first selected disorientations that provided comprehensive coverage of the disorientation space, though we did not attempt to use methods for uniform coverage, such as those described Quey et al.\ \cite{Quey:2018:uniformsampling}.

As noted in the introduction, computed GB structures are affected by the presence of free surfaces, which is why most datasets use periodic boundary conditions in the plane of the GB. This is readily solved when using CSL disorientations because the GB plane will be periodic for both grains if CSL lattice points are used when defining the supercell of the simulation. As such, all GBs used in this work are CSLs, but it should be noted that while low $\Sigma$ CSLs are typically considered to be special, there is evidence against associating CSLs with special properties \cite{Randle:1999vv,Randle:2006cj,Otto:2012:CSL_Creepresistance,EricksonHomer:2020hd}. Recent work showed that disorientation angle and BP are a better indicator of low GB energy than a low CSL $\Sigma$ value \cite{EricksonHomer:2020hd}. In other words, the CSL framework is used for its utility, not any attempt to sample ``special'' GBs.

The set of possible disorientations for this work was selected from a database of CSL values generated for $\Sigma$ values less than 1000; in all, this database contains 8554 CSLs and was generated following the methods of Grimmer \cite{Grimmer:1984ts}. 

To select a smaller set of CSLs, the list of 8554 CSLs was ordered from small to large $\Sigma$ values. Lower $\Sigma$ values have the advantage of smaller simulation supercells. The list of CSLs was then processed and each CSL was added to the list if it was not within $\Delta R$ of any CSL already on the list, otherwise it was discarded. $\Delta R$ in this case was defined based on an approximate $\sim5^\circ$ sampling, which in Rodriguez-Frank space is defined as $\Delta R=\tan(5^\circ/2) \approx 0.0437$. It is known that the distance between points changes in Rodriguez space on account of the tangent dependence of the rotation angle, but for simplicity a constant value of $\Delta R=\tan(5^\circ/2)$ is used throughout the disorientation fundamental zone.

\begin{figure}[t]
    \centering
    \includegraphics[width=\columnwidth]{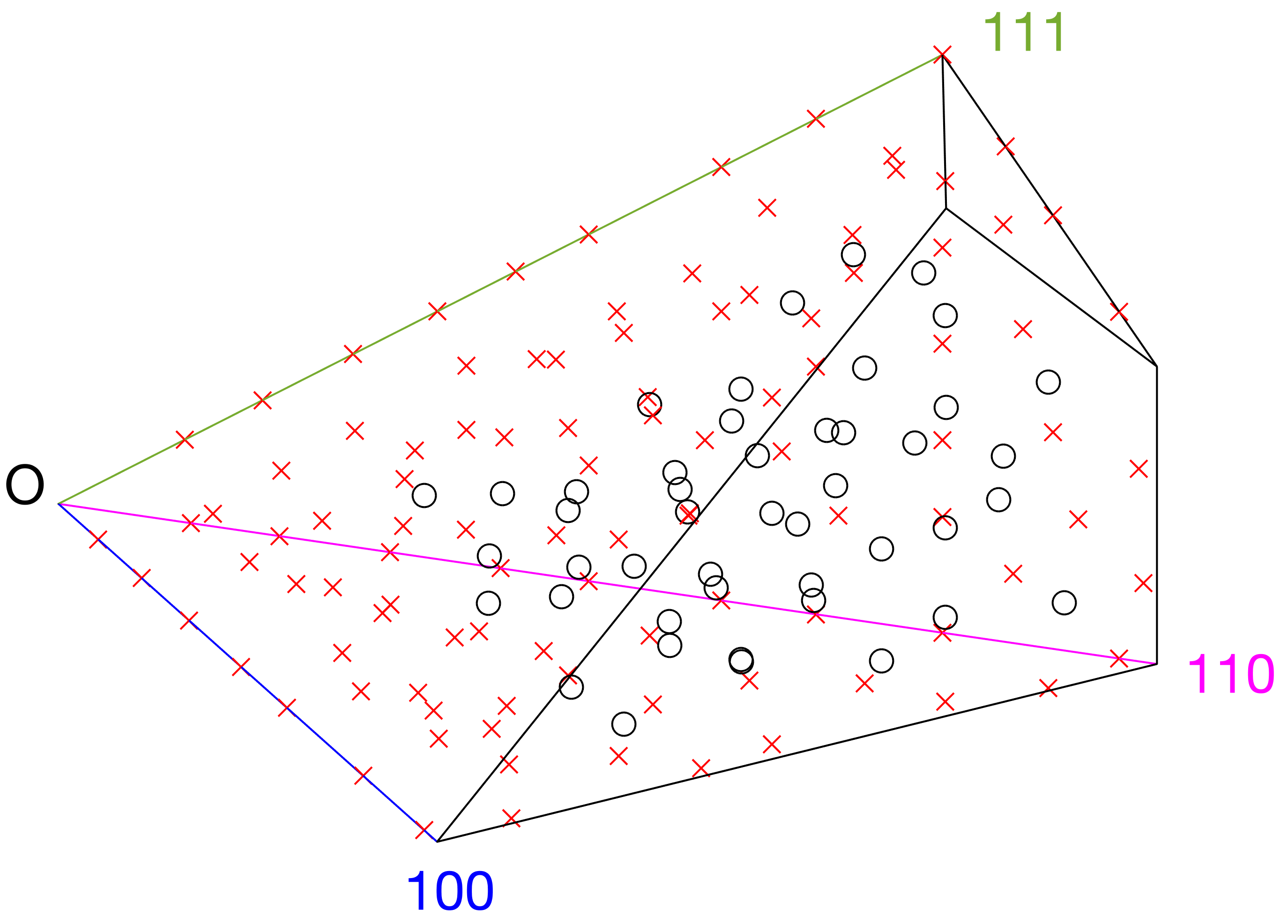}
    \caption{Plot of the 150 selected CSLs in the cubic-cubic fundamental zone in Rodriguez-Frank space. Points on the surface of the fundamental zone are marked with a red `x' while points on the interior are marked with a black `o'.}
    \label{fig:RFZ}
\end{figure}

Using the selection procedure described above, we obtained 150 CSLs. Of the CSLs in this set, 46 lie in the interior of the fundamental zone, 7 have 100 disorientation axes, and there are 9 on both the 110 and 111 disorientation axes. These 150 CSLs are plotted in the cubic-cubic disorientation fundamental zone in Rodriguez-Frank space in Figure \ref{fig:RFZ} and listed in supplemental Table S1. It is worth noting that there are 75 CSL with $\Sigma \le 99$, 35 CSLs with $99  < \Sigma \le 199$, 31 GBs with $199  < \Sigma \le 499$, and 9 GBs with $499  < \Sigma \le 999$; the largest of which is a $\Sigma999$ae.

From these 150 CSLs, we wanted to select GBs that would provide comprehensive coverage of the unique portions of the BP space for each CSL. The BP symmetries \cite{Patala.2013.PhilosophicalMagazine} enable the identification of the irreducible space or fundamental zone of unique BPs. The 3D point group symmetry of each CSL is listed in Supplemental Table S1. Table \ref{tab:BPstats} provides a summary of all the point group symmetries in this dataset along with the hemisphere coverage of the fundamental zone for that symmetry.

\begin{table}[t]
\centering
\caption{Statistics of the selected GBs and BPs, organized by the point group symmetry of the CSL lattice. \label{tab:BPstats}}
\begin{tabular}{ ccccc } 
 \hline
 \begin{tabular}[c]{@{}l@{}}Point \\ Group\end{tabular} & \begin{tabular}[c]{@{}l@{}}Hemisphere \\ Coverage\end{tabular} & \# CSLs & \begin{tabular}[c]{@{}l@{}}Average \# \\ of BPs/CSL \end{tabular}& \begin{tabular}[c]{@{}l@{}}Total \#\\ of BPs\end{tabular}\\
 \hline
     none &  1/1 & 46 & 54 & 2469\\
 $C_{2h}$ &  1/2 & 75 & 51 & 3834\\
 $D_{2h}$ &  1/4 & 13 & 40 & 525\\
 $D_{3d}$ &  1/6 &  8 & 30 & 240\\
 $D_{4h}$ &  1/8 &  7 & 32 & 223\\
 $D_{6h}$ & 1/12 &  1 & 13 & 13\\
 \hline
 & Total & 150 &  & 7304\\

 \hline
\end{tabular}
\end{table}

BPs were selected from each fundamental zone, with the goal of providing comprehensive coverage, such that one can interpolate energies over regions that are not sampled. This approach lead to much coarser spacing of points than occurred in the disorientation fundamental zone, but was necessary to keep the total number of GBs at a reasonable number.

\begin{figure}[t]
    \centering
    \includegraphics[width=\columnwidth]{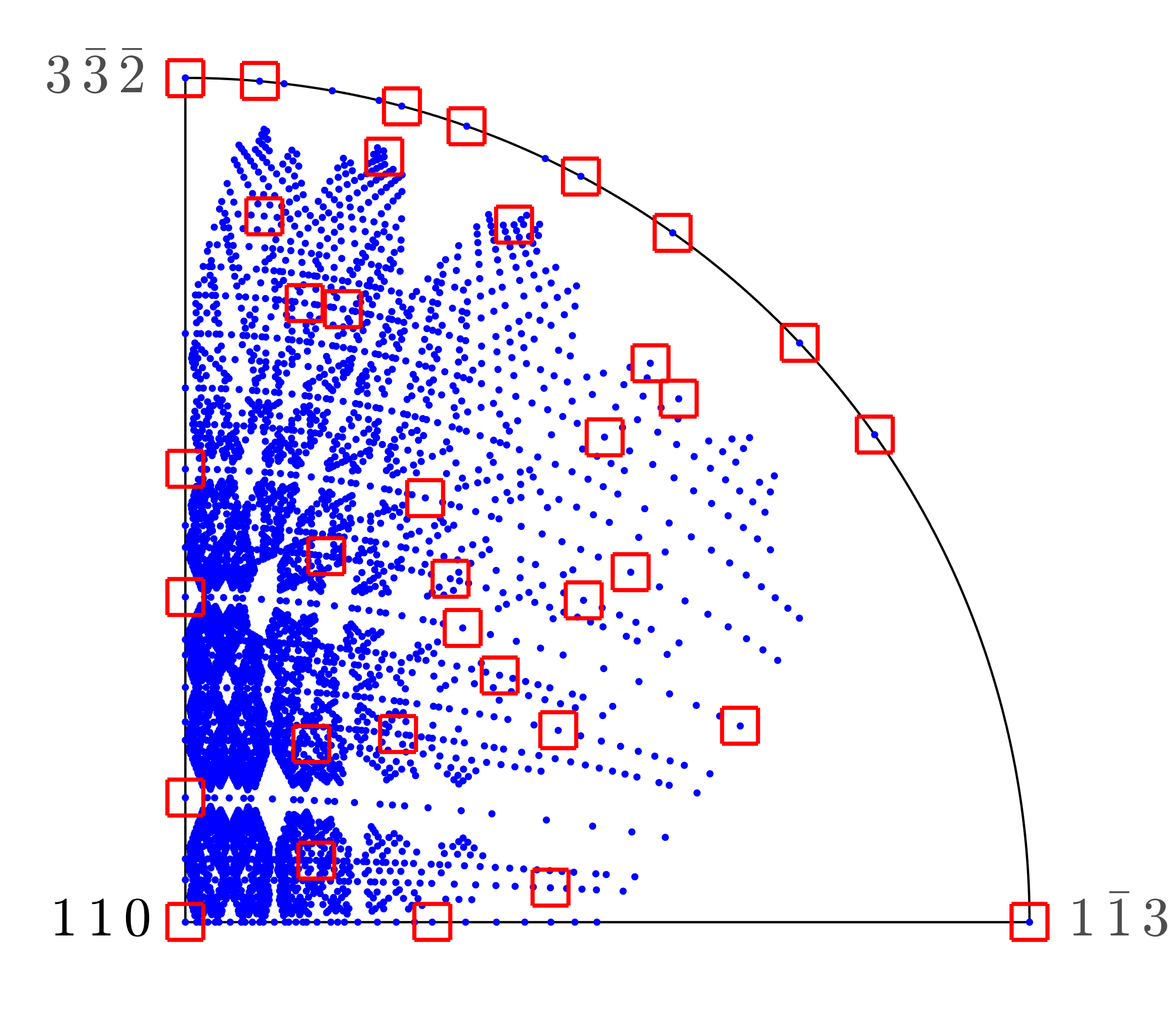}
    \caption{BP fundamental zone plot for the $\Sigma 9$ CSL marking possible BP orientations in blue that can be selected from $\pm4$ CSL lattice points and marking the selected BP orientations in red squares that aim to provide comprehensive coverage of the BP fundamental zone.}
    \label{fig:BPselectionS9}
\end{figure}

Unfortunately, the method used to select possible BPs does not lead to uniform coverage of each fundamental zone. This results from the use of a CSL lattice to select possible BPs, which will automatically satisfy periodic boundary conditions \cite{Homer:2015cv}. In order to generate possible BP normals that also contain CSL lattice points within the plane, the CSL lattice itself is used to create the possible BPs. It begins with the definition of several unit cells of the CSL lattice around the origin. Combinations of all pairs of CSL lattice points are used to define a possible BP, where the BP orientation is defined by the cross-product of the two CSL lattice vectors. These three vectors can then serve as the basis for a periodic GB supercell. Unfortunately, this approach restricts the possible BPs that emerge because the normals defined by the pairs of CSL lattice vectors do not point in all directions; the distribution is non-uniform. Fig.\ \ref{fig:BPselectionS9} illustrates the possible BPs, marked by blue dots, for the $\Sigma9$ CSL; it is clear that some BPs are difficult to access using this approach. The BPs for the sampling were automatically selected, followed by a manual selection of more BPs. The manual selection aimed to fill in gaps of the BP fundamental zone to enable more reliable interpretation of properties across the full fundamental zone. The BPs selected for the $\Sigma9$ CSL are marked with red circles in Fig.\ \ref{fig:BPselectionS9}.

A similar selection process of BPs was carried out for all 150 CSLs. The average number of BPs for each CSL is given in Table \ref{tab:BPstats} according to the point group symmetry; it is noted that those numbers do not provide consistent coverage in terms of the number of BPs per unit area of the BP fundamental zone. In the end, a total of 7304 GBs with unique disorientation and BP were selected for the dataset. It is worth noting that this is an order of magnitude greater in size than any other minimum GB energy dataset mentioned in the introduction, the dataset of 68,000 GBs created by Kim et al. are not necessarily representative of minimum energy GBs \cite{kim:2011}.

\subsection{Atomic GB Structure Creation}

Once selected, the vectors defining the basis for the periodic GB supercell are used to construct bicrystals. While many GB simulations require the basis vectors in the GB plane of the supercell to be orthogonal, we do not set this as a requirement because it allows a much larger population of BPs to be investigated \cite{Homer:2015cv}. However, in this work we choose to simulate a single BP, rather than two BPs, as is often done. This means that the simulation supercell is periodic in the GB plane, but not periodic in the direction normal to the GB. This also means that we do not have to define triclinic supercells, but rather monoclinic supercells. To account for the lack of periodicity, we follow standard approaches and place rigid blocks of atoms at the two extremes normal to the BP. These rigid blocks have atoms defined according to the lattice and the lattice constant \cite{Olmsted:2009ge,Tadmor:2011tb}. Thus, while these blocks have free surfaces, the atoms inside the crystal and adjacent to these blocks still experience the equivalent condition of being inside the bulk while also allowing the two crystals to translate. The minimum size for these supercells are similar to those defined by Olmsted et.\ al.\ \cite{Olmsted:2009ge} and defined as follows: at least $17a_o/2$ or two repeats of the CSL lattice in the direction of both basis vectors defining the BP and $20a_o$ plus an additional repeat of the CSL lattice unit cell in the direction normal to the BP on each side of the grain.

With the GB supercell dimensions defined, we consider the fact that to obtain structures that are likely to be the minimum energy GB structure, some form of search is required. While methods like USPEX and Monte Carlo searches exist for GBs \cite{Zhu:2017vna,banadaki:2018}, we choose to follow the methods of Olmsted and do a search over the various input variables to populating the GB supercell with initial atomic positions. Namely, we consider the variation of the following  six variables: translation of one grain relative to the other within the DSC lattice (3 variables), placement of the boundary interface (1 variable), allowed proximity of atoms before one is selected for deletion (1 variable), and three possible procedures for deleting overlapping atoms (1 variable). 

For the first 3 variables relating to grain translation, the DSC lattice is sampled at two points along each of the three basis vectors, resulting in 8 unique shifts of one grain relative to another. The placement of the boundary interface is carried out by searching for all unique combinations of plane spacings of the two crystals within the CSL unit cell; however, since plane spacings of high index planes can become quite small, a minimum step size is set at 0.1\AA. The allowed proximity of atoms before one is selected for deletion is set at 0.2\AA\ intervals between 33\% to 85\% of the nearest neighbor distance. Finally, when atoms are selected for deletion, the resulting structure is defined 3 different ways, one in which the selected atoms from the first grain are deleted, one in which the selected atoms from the second grain are deleted, and one in which the selected atoms from both grains are deleted and a new atom is placed at the average position of the two.

This variation of starting structures results in $6690$ possible starting structures on average, though the lowest number is 240 and the highest number is $101\,568$. It should be noted, however, that not all these starting structures are unique. Very often the variation of the parameters above results in an atomic structure that is identical to another. Once all possible structures are created, a 256 bit secure hash algorithm is run on batches of the structure input files to find and remove structures identical to one another. This reduces the number of structures that are required to 5888 on average, though the lowest number is 47 and the highest number is $78\,754$. Each of these structures is then minimized using conjugate gradient minimization in LAMMPS \cite{Plimpton:2022:LAMMPS}. In total, for all 7304 unique GBs, $43\,009\,236$ total structures have been minimized. After minimization, the GB energy is calculated following standard methods \cite{Tadmor:2011tb}. Note that structures are not thermalized in order to find lower minimum energy structures; the motivation behind not doing this is that thermalization often leads to faceting of the GBs, providing information about how that particular GB would like to evolve, but not providing a measure of energy for that particular BP orientation.

Saving the atom positions for all of these structures would have required too much storage , so the atoms beyond $\pm 15$\AA\ are deleted---only the atom positions near the GB are saved. All the structures for each unique GB are then compressed and saved in zip files for future examination. In total, the $43\,009\,236$ structures saved in zip files total more than 6 TB.

The dataset analyzed in this work focuses mostly on the minimum energy GB structure and corresponding properties of the 7304 unique GBs. These minimum energy GB structures are also analyzed for their GB width, based on the extent of atoms with non-FCC structure, as defined by common neighbor analysis (CNA), as well as for their excess volume per unit GB area, calculated according to standard methods \cite{Tadmor:2011tb}, using Python scripts in OVITO \cite{Stukowski:2010ky}. Data relevant to the GB disorientation, BP orientation, crystal orientations, minimum GB energy, number of structures minimized, structure file name, and excess volume, as well as all the minimum energy GB structures, are published in a dataset on Mendeley Data \cite{Homer:2022:AlGBdataset}.

\subsection{Machine Learning Methods}
\label{sec:methods:ML}
Characterizing microscopic degrees of freedom is a significant challenge, with considerable efforts focused on atomic structures generally \cite{Musil:2021,Bartok:2013cs,Ghiringhelli:2017,De:2016ia,Larsen:2016gt,Lazar:2018ji} and GBs specifically \cite{Snow.2019.StructureML,Patala:2019,Zhu:2017vna,Sharp:2018,Homer:2019fz,Rosenbrock:2018ug,Priedeman:2018gm,Rosenbrock:2017gl,Tamura:2017ja,Gomberg:2017ha,Zapiain:2020}. In this work no new structural characterizations or machine learning methods are employed. Rather, we repeat methods employed previously by some of the authors to demonstrate the utility and wealth of data found in this dataset as it pertains to machine learning methods.

As noted above, atoms within $\pm 15$\AA\ of the GB are available for analysis. In this work these atoms are analyzed using Smooth Overlap of Atomic Positions (SOAP) to characterize the local neighborhood of a selected atom \cite{Bartok:2013cs}. SOAP is a descriptor that encodes regions of atomic geometries by using a local expansion of a Gaussian smeared atomic density with orthonormal spherical harmonic and radial basis functions. In this work we employ the SOAP implementation in the QUIPPY package \cite{Csanyi2007-py,Bartok2010-pw,Kermode2020-wu}, which outputs a vector of all the coefficients for the basis functions. The SOAP parameters used are as follows: $r_\mathrm{cut}=3.74$, $n_\mathrm{max}=12$, $l_\mathrm{max}=12$, $\sigma=0.575$, $normalize=\textrm{true}$, and $Z=13$. It is also worth noting that we do not include atoms within $r_\mathrm{cut}$ of the ends of the simulation cells since their SOAP vectors do not match that of the bulk structure, despite their structure as bulk atoms.

It must be noted that feature matrices for machine learning must always be the same size and, for traditional machine learning methods, each data point must be represented by a single vector. The SOAP analysis exports a SOAP matrix for each GB, comprised of SOAP vectors for each atom, that can be different in size since each GB can have different numbers of atoms in the atomic structure file. To obtain features that are the same size and to reduce the feature representation to single vector, the SOAP vectors in this matrix are averaged to provide one average SOAP vector that represents the GB atomic structure as a whole, and is referred to as the average SOAP representation (ASR) in \cite{Rosenbrock:2017gl}.

While a matrix of these ASR vectors for each GB can be used as a feature matrix for machine learning, we choose to employ a kernel machine for the learning and must therefore construct a kernel from the ASR vectors. To be consistent with previous works \cite{Rosenbrock:2017gl}, we elect to define our kernel matrix as $AA^T$ where A is the ASR matrix. This results in a kernel matrix $K\in \mathbb{R}^{nxn}$ where $n$ is the number of features, 7304 in our case. The scikit-learn Python library is used to implement the support vector machine (SVM) algorithm \cite{scikit-learn}. The kernel from the ASR vectors essentially defines a similarity matrix between the GBs that is used by the SVM model to train and predict GB energy.

A second approach used for the machine learning in this work is referred to as the local environment representation (LER) \cite{Rosenbrock:2017gl}. The LER is a method to find a collection of ``unique'' local atomic environments by comparing the SOAP vectors of all GB atoms in the set and finding those that are unique as defined by a dissimilarity value $d$  less than a given magnitude $\epsilon$. The dissimilarity metric employed in this work is given by

\begin{equation}
\label{eq:similarity}
d_{\vec{a},\vec{b}}= \left|\left| \vec{a}-\vec{b} \right|\right|_2 = \sqrt{\vec{a}\cdot\vec{a} + \vec{b}\cdot\vec{b} -2\, \vec{a}\cdot\vec{b}}
\end{equation}

\noindent where $d_{\vec{a},\vec{b}}$ is the dissimilarity value when comparing the SOAP vectors $\vec{a}$ and $\vec{b}$ \cite{Priedeman:2018gm,Rosenbrock:2018ug}.

Once a set of these unique local atomic environments are found for all atoms $r_\mathrm{cut}$ away from the free surfaces in the set of 7304 GBs (50\,202\,187 atoms in total), the GB can then be represented by its fraction of each of these unique local atomic environments. The only parameter used to obtain the LER for the 7304 GBs in this work is $\epsilon = 0.1$. Using this parameter set, 101 unique local atomic environments were found and the resulting LER matrix gives the fraction of each of these unique environments for each GB as rows in the matrix. Similar to ASR, a kernel matrix is constructed as $LL^T$ where L is the LER matrix. The LER kernel matrix is then used as the feature matrix and GB energy is predicted using SVM after the same manner as ASR.

\section{Results \& Discussion}

\subsection{General Dataset Observations}

\begin{figure}[t]
    \centering
    \includegraphics[width=\columnwidth]{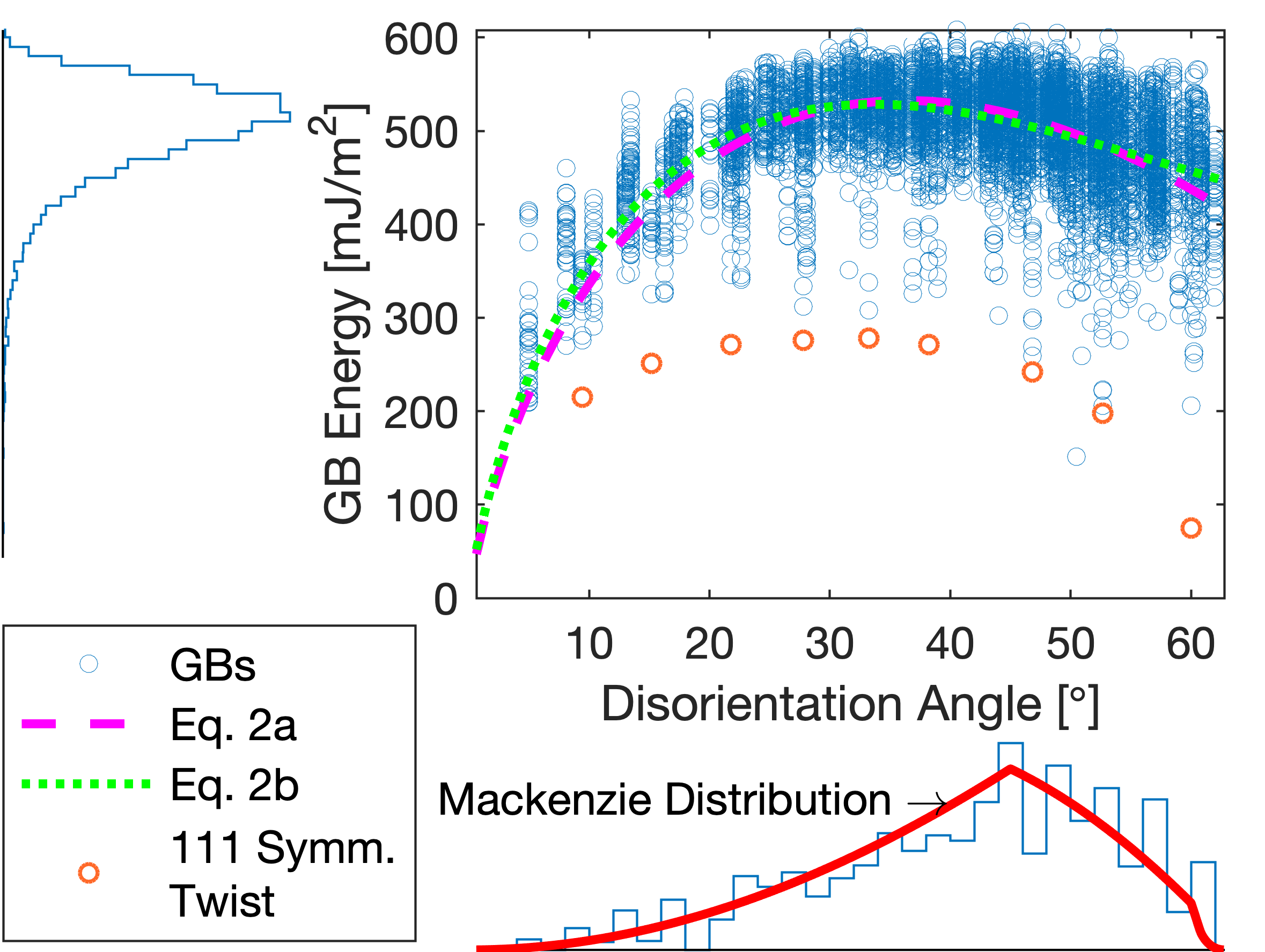}
    \caption{Scatter plot of energy as a function of disorientation angle for the 7304 GBs, accompanied by histograms of the two variables. Two possible Read-Shockley relationships are fit to the data and the low energy $1\,1\,1$ disorientation axis symmetric twist GBs are also identified.}
    \label{fig:energy_disangle}
\end{figure}

The simplest manner to illustrate all the data in a single plot is to plot the minimum GB energy values (referred to hereafter only as GB energy) as a function of a single variable, disorientation angle. This is shown in Fig.\ \ref{fig:energy_disangle} for all 7304 GBs. The scatter plot of this data is accompanied by histograms of GB energy and the disorientation angle.

The range of GB energies compares well with those reported for an Olmsted set of Al GBs \cite{Olmsted:2009ge}.
Note the strong unimodal peak in the GB energy distribution, which has mean and standard deviation values of  $497\pm55$ mJ/m\textsuperscript{2}. Note also the trend in the GB energy, which is similar to predictions made by Read-Shockley. In its most general sense, the Read-Shockley relationship can be defined as one of the following \cite{Wolf:1989ci}, where Eq.\ \ref{eq:RSlow} is valid for low angles and Eq.\ \ref{eq:RSall} is valid for all angles.

\begin{subequations}
\label{eq:RS}
\begin{align}
    & \gamma(\theta)_{low} = \theta\left[E_c - E_s \ln(\theta)\right]/b \label{eq:RSlow}\\
    & \gamma(\theta)_{all} = \sin(\theta)\left[E_c - E_s \ln\right(\sin(\theta)\left)\right]/b \label{eq:RSall}
\end{align}
\end{subequations}

The constants $E_c$ and $E_s$ represent the dislocation core energy and dislocation strain energy outside the core, respectively. These values are used as fitting parameters for the GB energy of the 7304 GBs in the dataset and their values are given in Table \ref{tab:RSfit}, along with the RMSE values from the fits. Fig.\ \ref{fig:energy_disangle} has the fitted Eqs.\ \ref{eq:RSlow} and \ref{eq:RSall} plotted on top of the individual data points. For reference, literature reports of dislocation core energy, $E_c$ are included \cite{Zhou.2017.DislocationCoreEnergy}, along with the range of $E_s$, based on screw and edge dislocations estimates, as given by Wolf \cite{Wolf:1989ci} from which this form of the equations are taken. Note that the fit values are near expectations even though individual dislocations are not identifiable in many of the high angle GBs where dislocations would overlap. 

It is noteworthy that an equation based on an idealized structure of dislocation arrays to mimic the structure and energy of GBs, developed over 70 years ago, captures the general behavior of a dataset that provides comprehensive coverage of the 5D space. While it is clear that these Read-Shockley fits do not describe all of the data points, the RMSE for both fits is 45 mJ/m\textsuperscript{2}. The utility of the Read-Shockley relationship is supported by other works as well, as indicated by Rohrer \cite{Rohrer:2011kp} and the fact that a modified form of Eq. \ref{eq:RSall} is used to interpolate GB energies between proximal boundaries in the Bulatov GB energy function \cite{Bulatov:2014bz}, which has now be used to describe energies of four fcc and two bcc metals \cite{Bulatov:2014bz,Ratanaphan:2021:IronGBE,Ratanaphan:2022:TungstenGBE}. Finally, it is worth noting that the Read-Shockley fits peak between 30-40\textdegree and higher disorientation angle GBs have slightly lower GB energies.

\begin{table}[t]

    \caption{Dislocation energy  parameters and RMSE values for the two Read-Shockley relationships, along with comparison of expected dislocation energies as taken from the literature.}
    \label{tab:RSfit}
\begin{tabular}{llll}
\hline
Source              & \begin{tabular}[c]{@{}l@{}} $E_c$ \\ $[\textrm{eV}/\textrm{\AA}]$ \end{tabular}  & \begin{tabular}[c]{@{}l@{}}$E_s$\\$[\textrm{eV}/\textrm{\AA}]$ \end{tabular}& \begin{tabular}[c]{@{}l@{}}RMSE\\ $[\textrm{mJ}/\textrm{m}^2]$ \end{tabular}\\ \hline
Eq.\ \ref{eq:RSlow} & 0.082        & 0.150         & 45 \\
Eq.\ \ref{eq:RSall} & 0.070        & 0.170         & 45 \\
Literature*        & 0.002-0.630   & 0.1893-0.2575 & -\\ \hline
\multicolumn{4}{l}{* Values for $E_c$ and $E_s$ are from \cite{Zhou.2017.DislocationCoreEnergy} and \cite{Wolf:1989ci}, respectively.}
\end{tabular}
\end{table}


The disorientation angle histogram in Fig.\ \ref{fig:energy_disangle} is compared with the Mackenzie Distribution, which represents the distribution of disorientation angles for a sample with a random distribution of cubic crystal orientations \cite{Mackenzie.1957.Biometrika,Mackenzie.1958.Biometrika}. The similarity between the distribution of disorientation angles in the dataset and the Mackenzie distribution are indicative that the sampling of GBs in this dataset is similar to what would be observed if a sample with random crystal orientations was analyzed. A histogram of just the 150 CSL disorientations is also provided in supplemental Figure S1.

To make use of this data for GB engineering, one must understand how to control the texture, or populations of different crystal orientations. From Fig.\ \ref{fig:energy_disangle} it is clear that a sample with random crystal orientations is unlikely to generate a GB distribution with many GBs that are low in energy. In fact, based on the distribution of energies in the histogram of Fig.\ \ref{fig:energy_disangle}, the fraction of low energy boundaries is bound to be negligible. Of course most GB engineering relies on the presence of twin boundaries and twin-related domains \cite{Randle:2004bz,Watanabe:2009jm,Randle:2010jw}, which are typically limited to low stacking fault energy FCC materials. Aluminum does not belong in this group, but research suggests that alignment of $\{1\,1\,1\}/\{1\,1\,1\}$ ``interconnected'' planes across GBs was frequently observed and has implications for GB engineering \cite{Wang:2018:MatChar}. The $\{1\,1\,1\}$ disorientation axis symmetric twist GBs, with $\{1\,1\,1\}/\{1\,1\,1\}$ alignment, define the low energy bound as a function of disorientation angle in Fig.\ \ref{fig:energy_disangle}, with the exception of the $\Sigma11$ $\{1\,1\,3\}/\{1\,1\,3\}$ symmetric tilt GB at a disorientation angle of 50.5$^{\circ}$. In other words, this data supports the appearance of these $\{1\,1\,1\}/\{1\,1\,1\}$ interconnected GBs, likely formed because of their low energy.

If we examine ``low'' energy GBs, defined by having an energy less than the highest $\{1\,1\,1\}$ disorientation axis symmetric twist GBs, we find 27 GBs from 6 CSLs. Three near $\{1\,1\,1\}/\{1\,1\,1\}$ GBs come from the $\Sigma131e \  (60.25^{\circ} \textrm{about } [5\,5\,4])$, which is vicinal to the $\Sigma3$, and therefore unsurprising to have low energy. 18 of these GBs come from the lowest disorientation angle GB studied in this work, the $\Sigma265a \  (4.98^{\circ} \textrm{about } [1\,0\,0])$, which is unsurprising considering low angle GBs are expected to have low energy. Another GB $\Sigma201a \  (8.09^{\circ} \textrm{about } [1\,1\,0])$, is also low angle, and therefore unsurprising to be included in this list. Finally, the $\Sigma11$ symmetric tilt GB, mentioned earlier, is well-known as a low energy GB. However, the remaining two GBs and CSLs are a bit surprising, these are a $\Sigma63c \  (54.03^{\circ} \textrm{about } [4\,3\,1])$ GB with BPs of $(5\,\bar{11}\,13)/(5\,\bar{1}\,\bar{17})$ and a $\Sigma69d \  (50.92^{\circ} \textrm{about } [5\,5\,1])$ GB with BPs of $(11\,\bar{1}\,19)/(1\,\bar{11}\,\bar{19})$. To the authors' knowledge, these particular GBs and CSLs have not previously been identified as having low energy. Due to the coarse nature of the sampling of the 5D space, there may also be other low energy cusps that exist but which have not been discovered.

In any case, the current dataset would suggest that beyond the three GBs mentioned above (from the $\Sigma11$, $\Sigma63c$, $\Sigma69d$ CSLs), an optimal texture would generate low angle GBs and $\{1\,1\,1\}/\{1\,1\,1\}$ interconnected GBs.

\begin{figure}[t]
    \centering
    \includegraphics[width=\columnwidth]{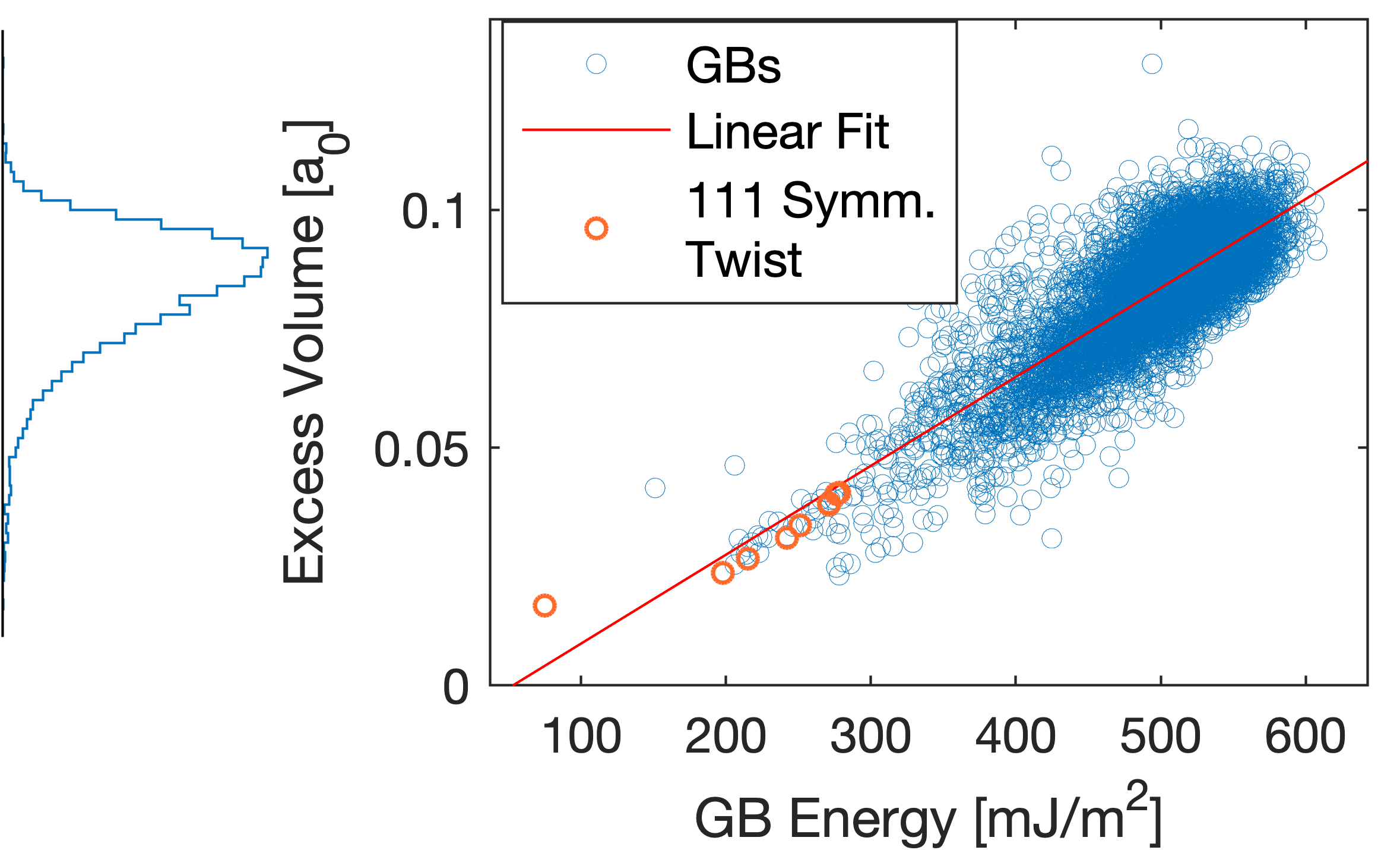}
    \caption{Scatter plot of excess volume, in units of the lattice parameter, as a function of GB energy for the 7304 GBs, accompanied by a histogram of the excess volume. The linear fit shown in the plot is given by $V_\mathrm{ex}=0.000187\times E_\mathrm{GB}-0.00989$.}
    \label{fig:gb_exvol}
\end{figure}

The excess volume per unit area, referred to here as the excess volume, is also calculated for each GB. The excess volume, normalized by the lattice parameter, is plotted against GB energy in Fig.\ \ref{fig:gb_exvol}, along with a distribution of the excess volumes. The range of GB excess volumes compares well with values of excess volume in an Olmsted set of Al GBs \cite{Olmsted:2009ge}. Note the strong correlation of excess volume with GB energy, which correlation has previously been shown to be linear \cite{Olmsted:2009ge}. The $\{1\,1\,1\}$ disorientation axis symmetric twist GBs are again highlighted and have both low GB energy and low excess volume, as would be expected based on the correlation.

\begin{figure}[t]
    \centering
    \includegraphics[width=\columnwidth]{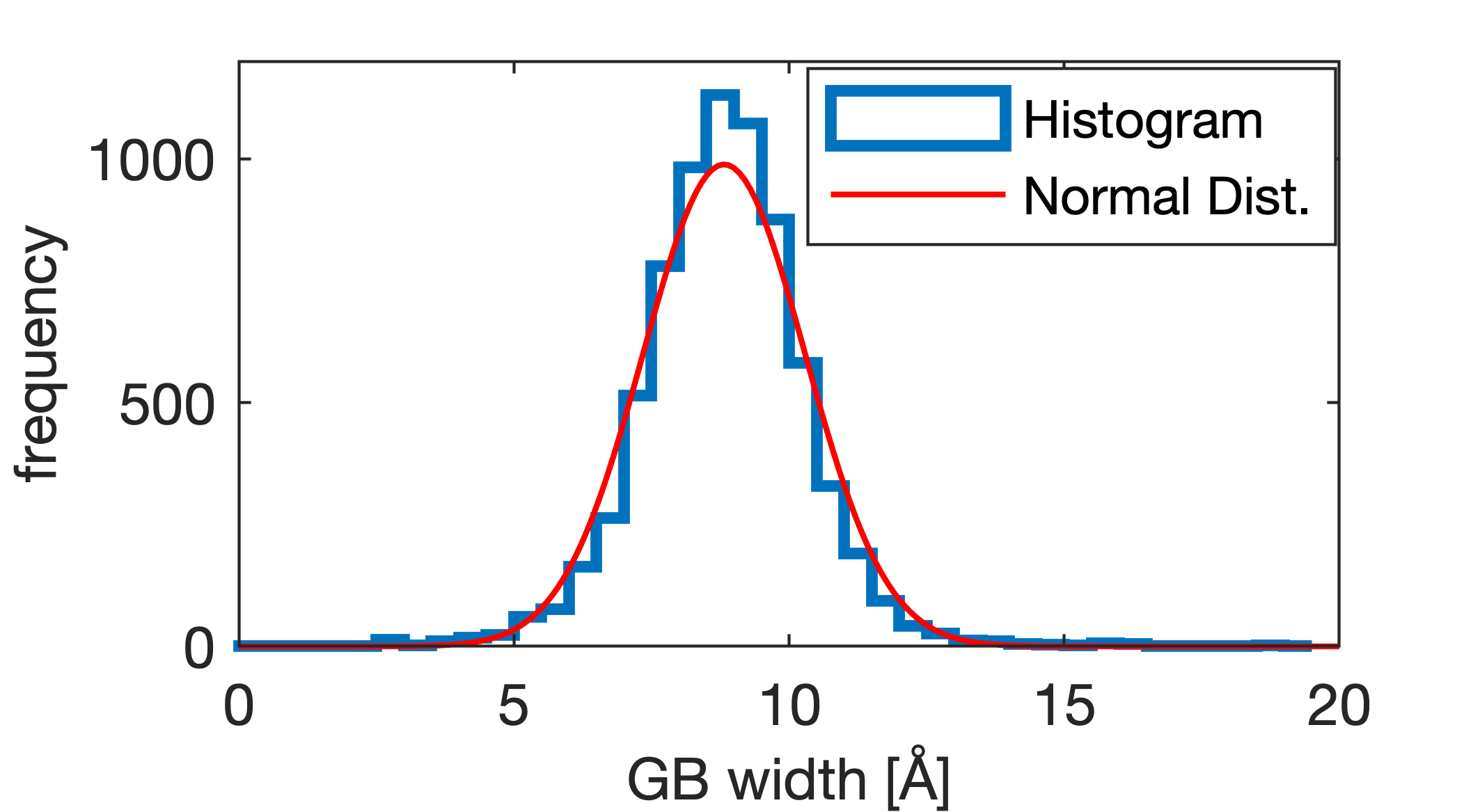}
    \caption{Histogram of GB widths compared with a normal distribution based on the mean and standard deviation of the data.}
    \label{fig:gb_width}
\end{figure}

Also of interest in this dataset is a chance to examine the distribution of widths observed in the GB structures. In this work, we define the GB width by finding the GB atoms, identified by CNA values that are not FCC, and then measuring the distance normal to the GB between the two most distant GB atoms on either side. Since it has been observed that defects, such as partial dislocations, can extend out from the GB some distance \cite{Homer:2013ce}, this width really measures the extent of the disorder. It does not mean that all the atoms within the ``GB width'' are disordered. The distribution of GB widths is plotted in Fig.\ \ref{fig:gb_width}, where the mean and standard deviation of the distribution are $8.8\pm1.5$ \AA. A Kolmogorov-Smirnov test indicates this distribution is normal with a significance value less than 0.001. These GB width values are in the range of estimates made by a variety of techniques in different materials
\cite{Zhu:1986:GBwidth,Wunderlich:1990:GBwidth,kuwano:1992:GBwidth,Page:2021jd}.

\subsection{GB energy trends in the 5D space}

\begin{figure*}[t]
    \centering
    \includegraphics[width=\textwidth]{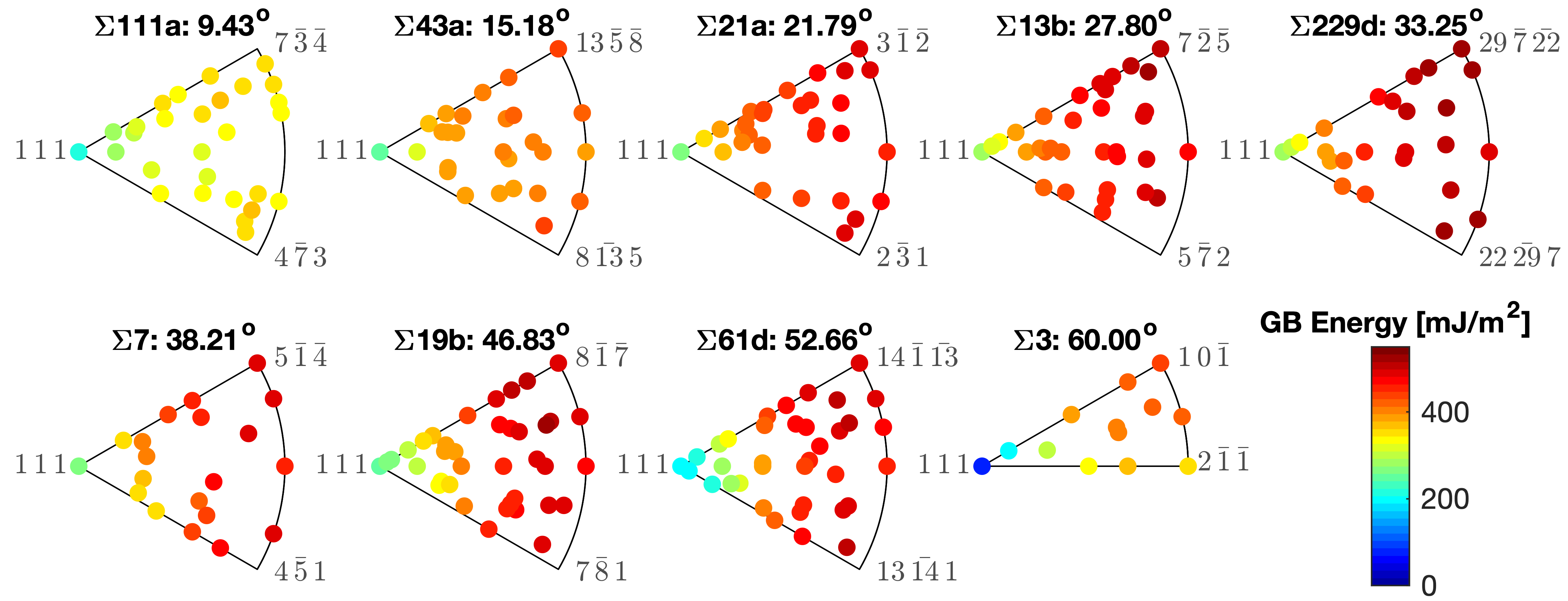}
    \caption{Plots of GB energy in the BP fundamental zones for 9 different $1\,1\,1$ disorientation axis CSLs. Note the similar trend of increasing energy from the $1\,1\,1$ symmetric twist GB toward the $1\,1\,1$ tilt GBs along the arc.}
    \label{fig:BPFZ111}
\end{figure*}

One of the major challenges in examining GB structure-property relationships is the fact that the crystallographic character is 5-dimensional. This becomes challenging to graphically illustrate, which is why properties are often plotted as a function of only one variable, as in Fig.\ \ref{fig:energy_disangle}. Here we seek to demonstrate that GB energy varies smoothly through this 5D space, which has been demonstrated in previous cases, often over limited scopes or with limited data \cite{Wolf:1989kv,Wolf:1989ts,Wolf:1990fk,Wolf:1990um,Saylor:2003ho,Saylor:2004bp,Rohrer:2011kp,Homer:2015ie,Bulatov:2014bz,Zhong:2017vd,EricksonHomer:2020hd,baird_homer_fullwood_johnson_2021}. Fig.\ \ref{fig:BPFZ111} plots the BP fundamental zones for several $1\,1\,1$ disorientation axis CSLs with increasing disorientation angle. Note the similarity in GB energy trends and the general increase in GB energy with increasing disorientation angle. Note also the fact that the $1\,1\,1$ symmetric twist GBs consistently have much lower energy values. In fact, these $1\,1\,1$ twist GBs are the lowest points in Fig.\ \ref{fig:energy_disangle}.

\begin{figure}[t]
    \centering
    \includegraphics[width=\columnwidth]{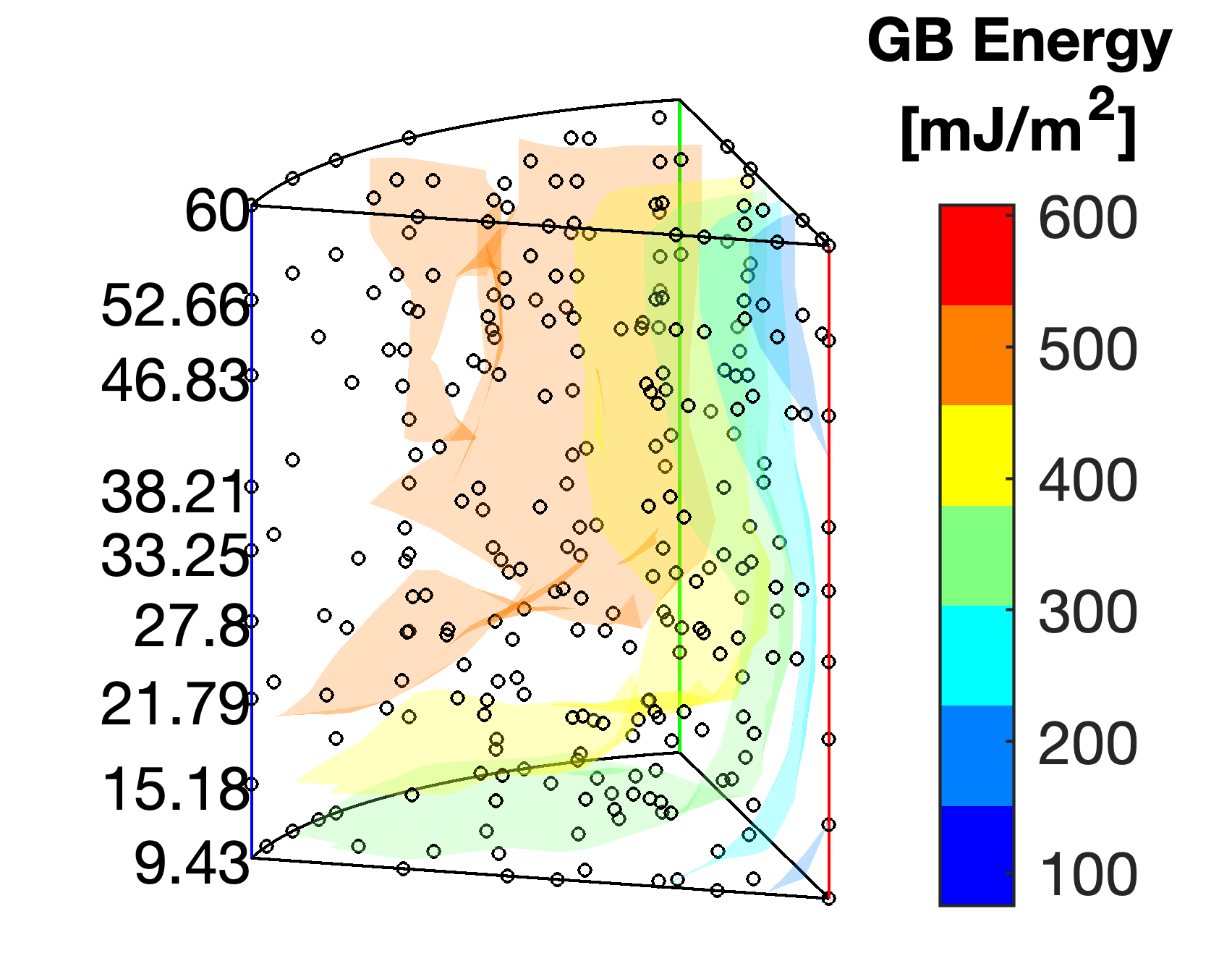}
    \caption{Disorientation angle-BP volumetric plot with isosurfaces of different GB energy values plotted to show the variation within the space of $1\,1\,1$ Disorientation axis GBs. This plot is essentially a vertical stacking of the images in Fig.\ \ref{fig:BPFZ111}.}
    \label{fig:cake111}
\end{figure}

Just as was done by Erickson and Homer \cite{EricksonHomer:2020hd}, these BP fundamental zones can be stacked as a function of disorientation angle to make a quasi-3D volumetric plot of energy. The stacking occurs such that the symmetric twist and tilt GBs, which are at the vertices of the plots in Fig.\ \ref{fig:BPFZ111}, are consistently also at the vertices of the 3D volume. This is illustrated in Fig.\ \ref{fig:cake111} where all the GB datapoints are marked with an `o' and isosurfaces of constant energy are included at various intervals to illustrate the gradient of energy between the sampled points. These isosurfaces illustrate how the energy increase as a function of disorientation angle is slower for the $1\,1\,1$ symmetric twist GBs than the rest of the boundary planes. It can also be seen that just like the Read-Shockley equation predicts, at the highest disorientation angles, the GB energy decreases in magnitude. 

While this work illustrates this variation along the $1\,1\,1$ disorientation axis, Erickson and Homer illustrated the variation along the $1\,0\,0$ disorientation axis in Ni \cite{EricksonHomer:2020hd}. These cases suggest that GB energy trends have similar trends for similar disorientation axes. This is supported by examinations of other disorientation axes where there were at least 3 different CSLs with the same disorientation axis. Supplemental Figs.\ S2-S5 show the isosurfaces of energy for these plots.

These evidences of smooth variation of GB energy in sections of the 5D space suggest that with the proper tools, one could examine trends of energy throughout the 5D space. Nearly all the 5D experimental datasets have examined GB energy throughout the 5D space and found repeated evidence that variation in BP orientation is associated with greater differences in GB energy than variation in disorientation \cite{Rohrer:2011kp}. This is supported by the present data as can be seen by the vertical lines of data points in Fig.\ \ref{fig:energy_disangle} and the variation of energy in Figs.\ \ref{fig:BPFZ111} and \ref{fig:cake111}. Additionally, Bulatov et al.\, in their GB energy function, note that the cusps and grooves follow the natural symmetries of the underlying lattice \cite{Bulatov:2014bz}. This can be seen in part by the fact that the low and high energy values in Figs.\ \ref{fig:BPFZ111} and \ref{fig:cake111} occur at the natural symmetry points or vertices of the BP fundamental zone. Some of these features can be seen in \cite{Homer:2015ie,EricksonHomer:2020hd} and in the supplemental Figs.\ S2-S5.

\begin{figure*}[t]
    \centering
    \includegraphics[width=\textwidth]{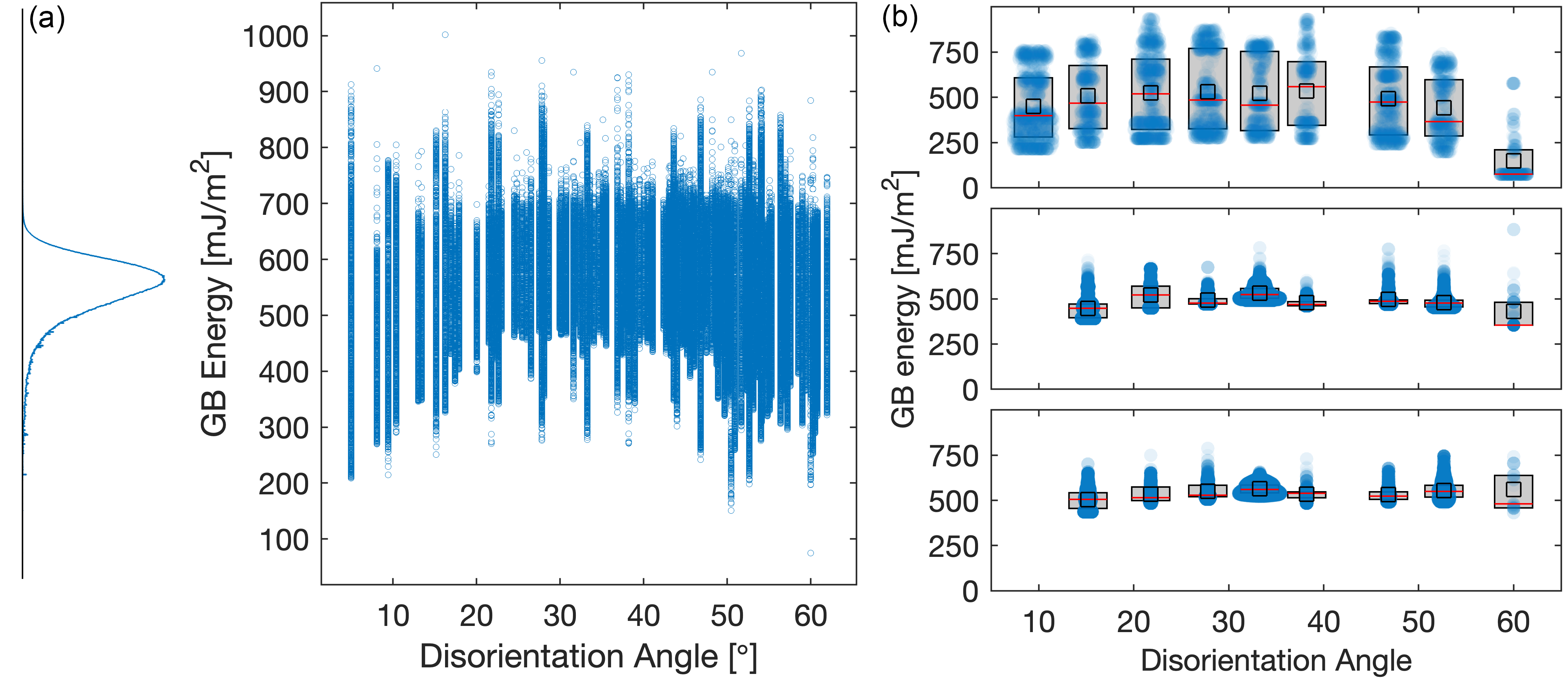}
    \caption{(a) Scatter plot of GB energy as a function of disorientation angle for all 43\,009\,236 minimized structures, accompanied by histograms of the GB energy. (b) Swarm plots of the GB energy for the $1\,1\,1$ symmetric twist, and the two $1\,1\,1$ symmetric tilt GBs from top to bottom, respectively. The swarm plots are overlaid with a box marking the first and third quartiles of the distribution, the red line marks the median, and the black square marks the mean.}
    \label{fig:metastability}
\end{figure*}

Recent developments in this area show great potential to understand this space further, with new distance metrics, interpolation schemes, expansions, visualizations, and other methods, to examine property GB character and property variations throughout the 5D space. For example, 
Francis et al. developed the octonion to compare and interpolate between GBs \cite{Francis.2019.Octonion}. This was followed by a number of tools to examine the GB manifold \cite{Chesser.2020.GBmanifold}. Baird et al.\ recently demonstrated an efficient interpolation technique in the 5D space using a Voronoi fundamental zone framework \cite{Baird:5DOFinterp} as well as methods for quantitative cartography of the GB energy landscape  \cite{baird_homer_fullwood_johnson_2021}. Hu et al.\ used a genetic algorithm and deep neural network to construct 5DOF property diagrams \cite{Hu.2020.GeneticML}. Mason and Patala recently introduced a new construction using orthornormal basis functions that account for crystallographic point group symmetries, grain exchange symmetry, and the null boundary singularity \cite{MasonPatala:2019:GB_BasisFunctions}. This sampling of recent works demonstrates that the community is poised to gain insight into the 5D space that was not possible before. It is hoped that the comprehensive 5D sampling provided by the present work will be a useful tool in gaining this new insight.

\subsection{Metastability}
As mentioned earlier, while producing this dataset, we stored the GB energy and structure for all structures that were minimized, not just the minimum energy case. In all, this amounted to 43\,009\,236 structures to obtain the minimum energy structures for the 7304 GBs. While these non-minimum energy structures are not included with the published dataset at this time, we present here a few insights into the energies of these additional structures. Fig.\ \ref{fig:metastability}a provides a plot of all 43\,009\,236 energies as a function of disorientation angle, along with a histogram of the energies. It can be seen that the distribution of energies in the histogram is not significantly altered from Fig.\ \ref{fig:energy_disangle}. The mean and standard deviation of the distribution are $546\pm55$ mJ/m\textsuperscript{2}. This is the same standard deviation but the mean is about 50 mJ/m\textsuperscript{2} higher.

Interestingly, it can be seen that the $1\,1\,1$ disorientation axis symmetric twist GBs that have the lowest energy are also the most likely to have the highest possible energy from one of their structures; this can be seen from the fact the fact that the vertical lines for these low energy GBs also extend well above the majority of boundaries in the energy distribution. We choose to analyze the symmetric twist and two symmetric twist GBs from the $1\,1\,1\,$ disorientation axis GBs, that are shown in Figs.\ \ref{fig:BPFZ111} and  \ref{fig:cake111}. In fact these correspond to the GBs whose normals are labeled at two of the three vertices as well as the mid-point of the arc. Swarm plot distributions for these  $1\,1\,1\,$ symmetric tilt and twist GBs are shown in Fig.\ \ref{fig:metastability}b. Here it is obvious that the symmetric twist GBs have both the lowest and highest GB energies. In contrast, the two symmetric tilt GB energies cover a much smaller range. For the whole dataset, the average and stand deviation of the range of energies covered by the structures for each of the 7304 GBs is $152\pm67$ mJ/m\textsuperscript{2}.

It is worth noting that work by Foley and Tucker examining damage tolerance of GBs follows a similar trend \cite{Tucker:2016:GBdamage}. In their work, the lowest energy GBs would absorb a significantly more defects and increase the free volume in the GB before achieving an equilibrium state at a much higher GB energy. In contrast, the GBs that started with a higher energy absorbed a smaller number of defects and increased the free volume in the GB by a smaller amount before achieving an equilibrium state at a modest increase to the GB energy. In short, the high energy GBs acted as more efficient sinks for defects, whereas the low energy GBs were not efficient sinks and accommodating defects resulted in large increases to the GB energy. If these $1\,1\,1$ disorientation axis symmetric twist GBs were to be advantageous because they have lower energy, in an environment where defect accumulation might be required (e.g., irradiation), the GB would quickly turn into a high energy GB, with possible energies in excess of that of the defect free high energy GBs, and maybe even in excess of their accumulated defect state.

It must also be mentioned at this point that we are unsure which of all these structures actually represent unique metastable states. For example, in Fig.\ \ref{fig:metastability}b it can be seen that many of the swarm plots have flat distributions at the bottom. This is representative of the fact that many of the starting configurations, although unique as an input file, found the same minimum energy configuration.

Previous methods to find metastable states \cite{Han:2016fi} used a finer sampling of 2D grain translations but did not consider other construction variables, such as boundary placement, which allowed them to determine which states were unique. In order to provide comprehensive coverage of the 5D space, we were not able to sample in such a manner. Methods such as USPEX \cite{Oganov:2006:USPEX,Oganov:2013:USPEX} are able to determine unique metastable states as the GB structure evolves \cite{Zhu:2017vna}. Zhu et al.\ used machine learning to determine the common structural features of GBs \cite{Zhu:2017vna} and such a method may prove fruitful in the future to determine which of all these structures represent true unique metastable structures.

If one can indeed determine which of all these structures represent unique metastable states, a number of potential research directions become available. First off, one can perform statistical mechanical calculations to determine expected values of different properties, as done by Han et al.\ \cite{Han:2016fi}. It is also possible that these metastable states represent the states through which a GB would pass as it migrates or deforms, as done by Alexander and Schuh \cite{Alexander:2013dk,Alexander:2016jv}. Such an approach would allow prediction of dynamic properties of a GB. Winter et al.\ recently devised a theory for nucleation of different GB phases \cite{Frolov:2022:GBnucleation}. Finally, with knowledge about the energy of different metastable states, and even better if they could be connected into a potential energy landscape, one could potentially predict thermodynamic properties, such as chemical potential, which is defined by the change in energy with respect to changes in the number of atoms. Thus, the metastable GB structures presented here have significant potential in advancing our understanding of GBs.


\subsection{Machine Learning}

\begin{table*}[!ht]
    \caption{Comparison of machine learning accuracy for ASR and LER as measured by RMSE and $R^2$ for a variety of training and validation splits of the dataset. For the answers with 5-Fold cross-validation (CV), the RSME and $R^2$ are given as the mean $\pm$ the standard deviation. The low and high disorientation angle split is defined at a disorientation angle of  15$^{\circ}$, with 250 and 7054 GBs in the groups, respectively. }
    \label{tab:ML}
\centering
\begin{tabular}{l c c c c}
 \hline
    
   Characterization and Dataset Split & \multicolumn{2}{c}{RMSE [mJ/m\textsuperscript{2}]} & \multicolumn{2}{c}{$R^2$}\\
   \hline
   ASR  & Train & Validation  & Train & Validation \\ 
 \hline
 5-Fold CV Random Split & 12.85$\pm$0.09  & 12.88$\pm$0.33 & .945$\pm$0.001  & .945$\pm$0.004\\ 
 5-fold CV Disjoint Disorientation Angle Split & 10.45$\pm$1.10 & 13.17$\pm$1.15 & .94$\pm$.03 & .94$\pm$.01\\
 Disorientation Angle Train Low-Validation High  & 13.38 & 30.35 & .96 & .63\\
 Disorientation Angle Train High-Validation Low  & 9.21 & 11.04 & .97 & .97 \\
 Shuffled GB energies Train 2/3-Validation 1/3 & 56.58 & 54.64 & -.037 & -.036 \\
\hline
   LER & Train & Validation   & Train & Validation\\ 
 \hline
 5-Fold CV Random Split & 19.73$\pm$0.13  & 20.04$\pm$0.52 & .87$\pm$0.002  & .87$\pm$0.007\\ 
 5-fold CV Disjoint Disorientation Angle Split & 17.53$\pm$1.93 & 23.68$\pm$1.35 & .86$\pm$.051 & .81$\pm$.042\\
 Disorientation Angle Train Low-Validation High  & 14.47 & 44.19  & .95 & .23 \\
 Disorientation Angle Train High-Validation Low  & 19.42 & 26.27  & .85 & .85 \\
 Shuffled GB energies Train 2/3-Validation 1/3 & 56.35 & 54.81  & -.025 & -.042 \\
 \hline
\end{tabular}
\end{table*}

\begin{figure}[t]
    \centering
    \includegraphics[width=\columnwidth]{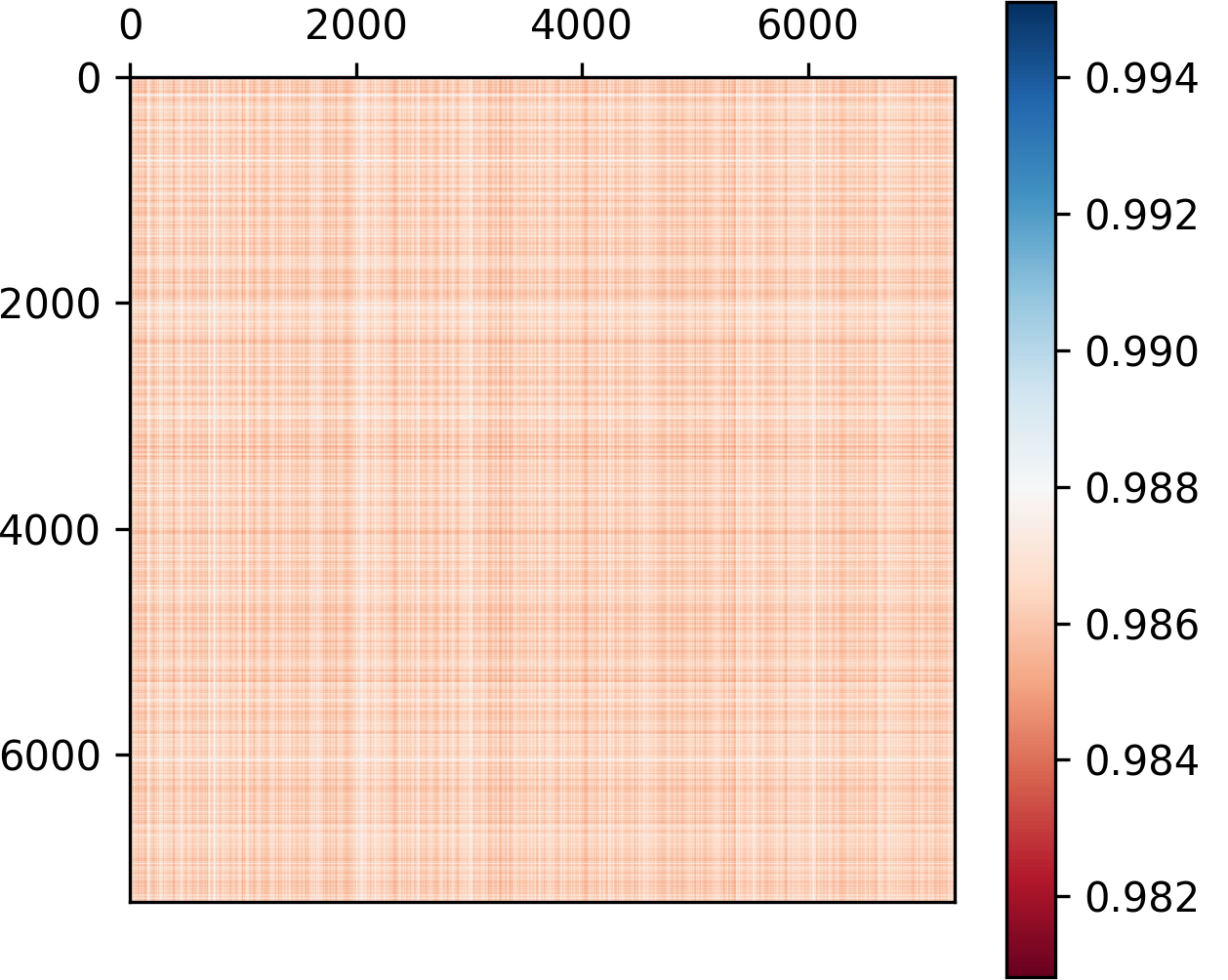}
    \caption{ASR kernel matrix representing the similarity between all 7304 GBs.}
    \label{fig:kernel}
\end{figure}

\begin{figure*}[th!]
    \centering
    \includegraphics[width=\textwidth]{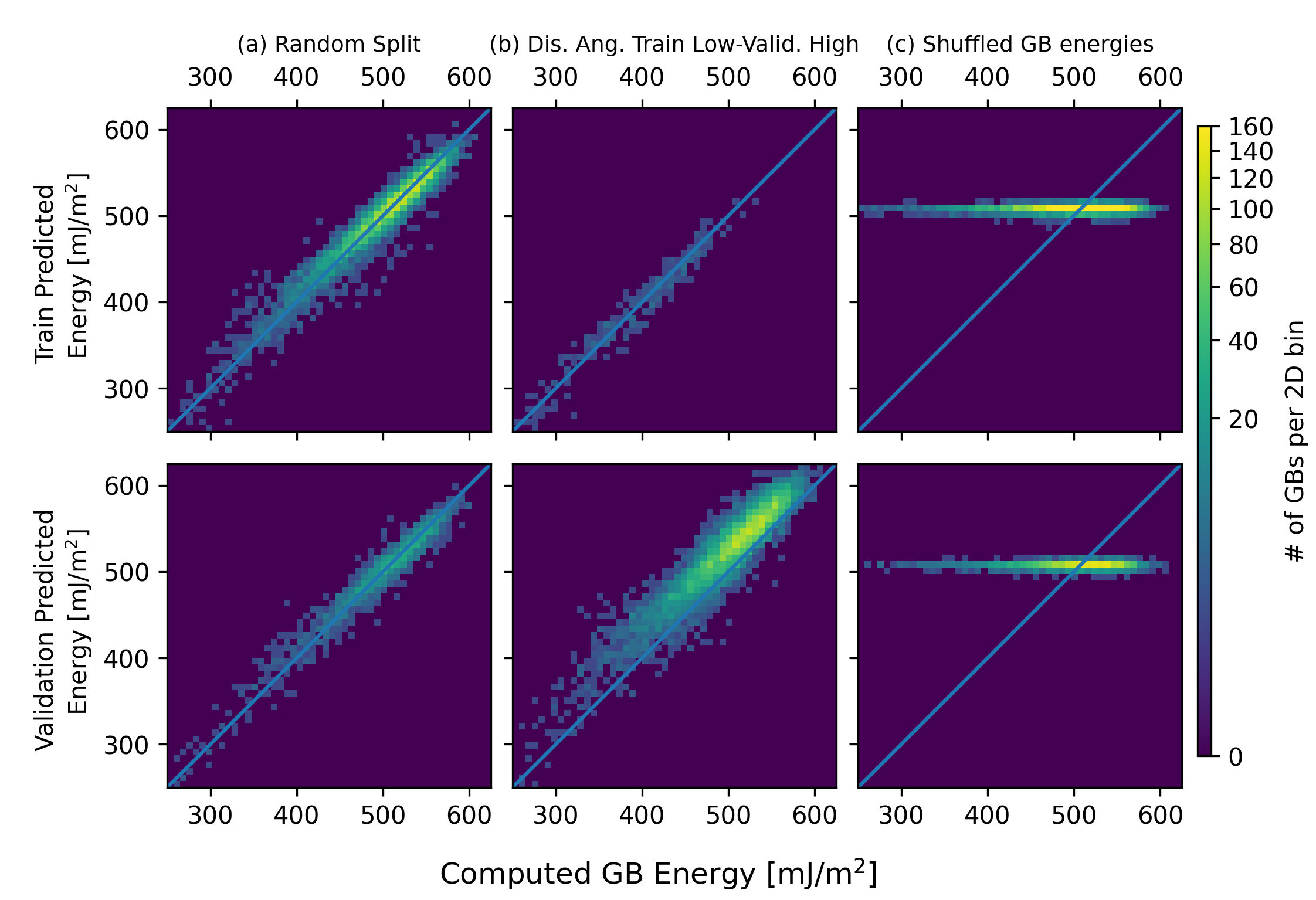}
    \caption{Parity plots comparing predicted vs.\ computed GB energy for both the training and validation sets trained on ASR. The various splits match some of those described in Table \ref{tab:ML}, and are identified as (a) one of the 5-Fold cross-validation random splits, (b) the disorientation angle train low-validation high split, and (c) the shuffled GB energies train 2/3-validation 1/3 split. }
    \label{fig:ML}
\end{figure*}

Here we turn our attention to the atomic structure of the GBs, which represent the microscopic degrees of freedom of a GB. As described in the methods (\ref{sec:methods:ML}), we characterize all the atomic structures of the 7304 GBs using both ASR and LER. Also mentioned in the methods is that, in this work, we make no effort to innovate in characterization or machine learning approaches. Rather, the focus here is to illustrate the richness of this dataset as it pertains to potential uses in machine learning.

As noted in the methods, following ASR characterization, a kernel matrix is defined that represents the similarity between all the GBs. This kernel matrix is illustrated in Fig.\ \ref{fig:kernel} where it can be seen that the the range of values in the kernel is quite small. Clearly some GBs are more like and unlike other GBs, but the interesting thing to note here is that this feature matrix does not appear to provide significant discrimination.

With the feature matrix defined by the kernel matrix, we run support vector regression with various splits of the data into training and validation sets. These splits are defined in Table \ref{tab:ML}. In the first split, a 5-fold cross-validation is carried out with a random split of the GBs. Since a random split could allow training of the model at nearby points, we perform a 5-fold split that is disjoint over the disorientation angles, ensuring that each split has approximately the same number of boundaries. We also split on disorientation angle into low and high angle GBs, with the split at $15^{\circ}$. We perform two tests, training on the low angle GBs and validation on the high angle GBs and vice versa. Finally, for a control, we perform a random shuffling of the GB energy values and train and validate the model. 

The results of all these tests are shown in Table \ref{tab:ML} as measured by the RMSE and $R^2$ values. This is accompanied by Fig.\ \ref{fig:ML} which has a 2D visualizations comparing the computed and training predictions of GB energy and the computed and validation predictions of GB energy for 3 of the splits listed in Table \ref{tab:ML}. In Fig.\ \ref{fig:ML}, a perfect prediction either in the training or validation would fall along the parity line. The first column has one of the 5 folds in the random split for the cross validation, where the predictions are good as evidenced by both the RMSE and $R^2$ values. The disjoint disorientation angle split also has low RMSE and high $R^2$ values. 

The train low-validate high disorientation angle split is illustrated in the second column where only 250 low angle GBs are used to train the model and predictions are made on the 7054 validation GBs. The validation RMSE value is nearly double the previous examples and the $R^2$ value is considerably lower, but still indicative that more than half of the variance in the predicted energy is accounted for in the model. It is clear however that the machine learning model trained on the low angle GBs systematically underpredicts the energy for the high angle GBs, which accounts for much of the changes in RMSE and $R^2$ values. 

Finally, the last column in Fig.\ \ref{fig:ML} has a random shuffling of the GB energy values such that there should be no correlation to be found in the machine learning. In the training and validation, the $R^2$ value near zero is indicative that the machine learning finds no correlation, as expected. This is clearly evident in the figure, and the predictions are just over the mean value of the GB energy shown in Fig.\ \ref{fig:energy_disangle} 497 mJ/m\textsuperscript{2} and the RMSE value is about equal to the standard deviation in Fig.\ \ref{fig:energy_disangle} of 55 mJ/m\textsuperscript{2} for both the training and validation. Thus, the random shuffle is only able to predict the mean, but the large concentration of the GB energy about the mean value means that the RMSE values are low. This also means that we must take care in interpreting the RMSE values because of the distribution of energy values.


The same machine learning process is repeated with the LER kernel matrix and, similar to previous works \cite{Rosenbrock:2017gl}, the LER learning is not as good as the ASR learning, as evidenced by the RMSE and $R^2$ values in Table \ref{tab:ML}. While the initial goal of LER was to get away from the averaging and information loss that occurs in ASR, it is notable that the predictions from ASR are consistently better. The reason for this may be that while LER attempts to get away from a single ``average'' description in favor of representing GB structure from a distribution of unique local atomic environments, it discards all information about those unique atomic environments calculated with SOAP. This is then reduced further in the formation of the kernel, thus losing additional information.

To reiterate, the goal of showing this learning is to demonstrate that this dataset will be useful for machine learning approaches to find structure property relationships of GBs that can focus on the microscopic (atomic) degrees of freedom. So, while these two approaches provide noteworthy learning and validation, the challenge remains to find feature representations that are consistent in size and also encode relevant information about the GB structure for learning on a variety of properties and phenomena. We are optimistic that this dataset will provide sufficient data for individuals to devise and test clever characterization and machine learning approaches to advance GB structure-property relationships that span the full 5D crystallographic (macroscopic) character while taking into account the atomic (microscopic) degrees of freedom that define the structure and properties of GBs.

\section{Conclusions}

This work examines a computed dataset of aluminum GBs that provide comprehensive coverage of the 5D crystallographic character space. The dataset is comprised of 7304 GBs that cover a range of BPs in 150 different CSLs that span disorientation space. For each GB, a number of possible structural configurations is considered and minimized. This results in more than 43 million possible structures that could be metastable states corresponding to the 7304 GBs. The following observations are made concerning this dataset, with a particular focus on the minimum energy GB structures and properties.
\begin{itemize}
\item The distribution of disorientation angles in the dataset match a Mackenzie distribution, indicating equivalence to GBs that would be present in a cubic sample with random crystal orientations.

\item The distribution of GB energies in the dataset is unimodal with a mean and standard deviation of $497\pm55$ mJ/m\textsuperscript{2}.

\item The GB energies as a function of disorientation angle follow the general trend of the Read-Shockley relationship with a peak between 30-40\textdegree, with a subsequent decrease in energy for the highest disorientation angles.

\item The $\{1\,1\,1\}/\{1\,1\,1\}$ interconnected GBs have the lowest GB energies for all disorientation angles, with the exception of the $\Sigma11$. This is consistent with experimental observations where these $\{1\,1\,1\}/\{1\,1\,1\}$ interconnected GBs are expected to have low energy. 

\item The GB energy and excess volume per unit area show a strong linear correlation.

\item The GB width, as measured by the extent of disorder normal to the GB, has a normal distribution with a mean and standard deviation of $8.8\pm1.5$ \AA.

\item The GB energy, as examined in subspaces of the 5D space, show smooth variation and similar behaviors for similar disorientation axes.

\item Examination of all the GB energy values obtained during the sampling of the 43\,009\,236 structures is also Read-Shockley-like, and the distribution of values is unimodal with a mean and standard deviation of $546\pm55$ mJ/m\textsuperscript{2}.

\item Examination of all the GB energy values obtained during the sampling of the $1\,1\,1$ disorientation axis symmetric twist GBs, which are the same $\{1\,1\,1\}/\{1\,1\,1\}$ interconnected GBs, have the largest spread in GB energies for the non-minimum energy structures, ranging from the lowest possible energies at a given disorientation angle to the highest possible GB energies. In contrast the $1\,1\,1$ disorientation axis symmetric tilt GBs, which have higher GB energy, have a much smaller range. This is consistent with observations of damage accumulation where low energy GBs are unable to accommodate defects without a significant increase in GB energy.

\item Machine learning of GB atomic structures to predict GB energy using previously published methods, referred to as average SOAP representation (ASR) and local environment representation (LER), perform well on training and validation sets as measured by RMSE and $R^2$ values. This is true even when training on small datasets, such as training on 250 low angle GBs and predicting on all 7054 high angle GBs.

\end{itemize}

The authors are optimistic that this dataset will be useful for learning GB structure-property relationships on both the 5D crystallographic (macroscopic) character as well as the atomic (microscopic) structure of GBs. As an example, a large database of handwritten numbers, known as MNIST, has long provided fertile ground for training of image processing methods and other machine learning models and serves as a basis for benchmarking different algorithms \cite{lecun-98,Deng:2012:MNIST,Bengio:2013:MLreview_MNIST}. With the explosion of methods to examine GBs at both the macroscopic and microscopic scales, this dataset may prove useful in a similar way.


\section*{Dataset}
The dataset for this work is published on Mendeley Data \cite{Homer:2022:AlGBdataset}, with an embargo date of 12/31/2022. Individuals interested in obtaining the dataset prior to this date may contact the authors. The dataset includes a CSV file with details about the crystallographic character, calculated properties, and other information about the 7304 GBs in the dataset. The CSV file is accompanied by a zipped directory containing the atomic structures of all the minimum energy configurations for the 7304 GBs in this work.

\section*{Acknowledgments}
This work was supported by the U.S.\ National Science Foundation (NSF) under Award \#DMR-1817321.

\bibliographystyle{elsarticle-num}
\bibliography{Research_Paper_Bib}
\end{document}